# Interfacial and Surface Magnetism in Epitaxial NiCo$_2$O$_4$(001)/MgAl$_2$O$_4$ Films

Corbyn Mellinger,[1] Xiao Wang,[2] Arjun Subedi,[1] Andy T. Clark,[2] Takashi Komesu,[1] Richard Rosenberg,[3] Peter A Dowben,[1,4] Xuemei Cheng,[2] Xiaoshan Xu[1,4*]

[1] *Department of Physics and Astronomy, University of Nebraska, Lincoln, NE 68588, USA*

[2] *Department of Physics, Bryn Mawr College, Bryn Mawr, PA 19010, USA*

[3] *Advanced Photon Source, Argonne National Laboratory, Lemont, IL 60439, USA*

[4] *Nebraska Center for Materials and Nanoscience, University of Nebraska, Lincoln, NE 68588, USA*

**Abstract**

NiCo$_2$O$_4$ (NCO) films grown on MgAl$_2$O$_4$ (001) substrates have been studied using magnetometry, x-ray magnetic circular dichroism (XMCD) based on x-ray absorption spectroscopy, and spin-polarized inverse photoemission spectroscopy (SPIPES) with various thickness down to 1.6 nm. The magnetic behavior can be understood in terms of a layer of optimal NCO and an interfacial layer (1.2± 0.1 nm), with a small canting of magnetization at the surface. The thickness dependence of the optimal layer can be described by the finite-scaling theory with a critical exponent consistent with the high perpendicular magnetic anisotropy. The interfacial layer couples antiferromagnetically to the optimal layer, generating exchange-spring styled magnetic hysteresis in the thinnest films. The non-optimal and measurement-speed-dependent magnetic properties of the interfacial layer suggest substantial interfacial diffusion.

## I. Introduction

The spinel material NiCo$_2$O$_4$ (NCO) has seen a dramatic increase in research interest recently due to its applications in energy sciences [1–3] and catalysis [4,5], with particular interest in the effect of nanoscale structures. More fundamentally, the high conductivity, high perpendicular magnetic anisotropy (PMA), and high magnetic transition temperature [6–9] of NCO thin films befit data storage applications.

The significant magnetization (~2 $\mu_B$/f.u.) and anisotropy of NCO is known to be driven by the specific cation stoichiometry and occupation on the tetrahedral ($T_d$) and octahedral ($O_h$) sites [Fig. 1(a)]. In particular, the $T_d$ sites are occupied by the high-spin Co$^{3+}$ and Co$^{2+}$ ions; the 50% of the octahedral ($O_h$) sites are occupied by the non-magnetic Co$^{3+}$ ions; the other 50% of the $O_h$ sites are occupied by the Ni$^{3+}$ and Ni$^{2+}$ ions whose magnetic moments anti-align with the Co moments [9,10]. Indeed, several studies focused on the growth conditions of NCO by modulating growth pressure [7,8], growth temperature [10], and deposition laser rate [11] have concluded optimal magnetic properties to be due to the site occupations and stoichiometry outlined above [12].

More recently, the effects of film thickness as it relates to magnetotransport and magnetic orders have been studied. The results show decrease of magnetic transition temperature and magnetization and reveal a transition to insulating state in ultrathin films [13,14]. Understanding the behavior of the ultrathin films, the interfacial, and surface properties is essential, as work is already being produced using NCO in spintronic applications, such as within a magnetic tunnel junction [15].

To elucidate the interfacial and surface magnetism of the NCO films, we have prepared NCO films with a variety of thicknesses and studied them using magnetometry, x-ray magnetic circular dichroism (XMCD) spectroscopy, and spin polarized inverse photoemission spectroscopy (SPIPES). Our results indicate an interfacial layer of non-optimal cation stoichiometry (≈ 1.2 nm) that prefers antiferromagnetic alignment with the optimal part of the film. A small in-plane magnetic moment is observed for the surface layer, suggesting a slight canting.

## II. Methods

NiCo$_2$O$_4$ (001) films were grown using pulsed laser deposition on MgAl$_2$O$_4$ (001) (MAO) substrates at 350˚C, with a KrF excimer laser of 6 Hz repetition rate, 80 mJ pulse energy, and background oxygen pressure $P_{O2}$ = 150 mTorr. Magnetometry measurements were performed using a superconducting quantum interference device (SQUID) with temperatures up to 400 K and applied fields up to 60 kOe. X-ray diffraction (θ-2θ) and reflectivity (XRR) were done using a Rigaku SmartLab system with a Cu-Kα source (λ = 1.54 Å) [Fig. S1].

X-ray absorption (XAS) and X-ray magnetic circular dichroism (XMCD) measurements were performed at Beamline 4-ID-C of the Advanced Photon Source at Argonne National Laboratory. Samples were measured with magnetic fields up to ±2 kOe applied perpendicular to the film plane at temperatures between 80 and 320 K. XAS was performed at the Co and Ni L$_{2,3}$ absorption edges in both total electron yield (TEY) and total fluorescence (TFY) modes. Magnetic hysteresis loops were measured by sweeping the magnetic field at the Co and Ni L$_3$ absorption edges, 777.0 eV and 851.7 eV, respectively. The XMCD contrast signals are converted to magnetization using the XMCD sum rules by scaling with the XAS spectra [16]. SPIPES of 12 nm thick NCO/MAO was carried out at 183 K with a base pressure better than $8 \times 10^{-10}$ torr, as described elsewhere [17].

## III. Results

### A. Magnetometry

Magnetization vs. temperature curves measured using magnetometry between 20 and 400 K are shown in Fig. 1(b). Prior work has indicated that high $O_2$ back ground pressure ($P_{O2} > 100$ mTorr) results in optimal (high) Curie temperature $T_C$ and saturation magnetization $M_s$ [18]. Additional work has demonstrated that reduced growth rate increases $T_C$ to higher than 400 K [19]. Consistent with these studies, the results in Fig. 1(b) indicate these NCO films were grown under the optimal condition [12].

Overall, the measured magnetization of the NCO films increases with thickness and appears to saturate in thick films. In Fig. 1(c), total moments at 20 K are plotted against the film thickness. The relation can be well fit using a linear function $\mu = M_{bulk}(t - t_0)A$, where μ is the total moment, $M_{bulk}$ is the bulk saturation magnetization, $t$ and $A$ are the film thickness and area respectively, $t_0$ is the intercept on the horizontal axis. The fit results are $t_0 = 1.2 \pm 0.1$ nm and $M_{bulk} = 2.1 \pm 0.1$ $\mu_B$/f.u. While the $M_{bulk}$ value agrees well with the previous results for optimal growth condition [12], the finite $t_0$ value suggests the presence of an interfacial layer, which is similar with the observation in other magnetic thin films [20–24]. Previously, NCO films grown in lower $P_{O2}$ were reported to have a thicker (1.6 nm) interfacial layer [24], indicating the impact of the growth condition on the interface.

As shown in Fig. 1(d), the magnetic transition temperature ($T_C$) also reduces in the thinner films, which can be understood using the finite scaling theory [25–28]. For bulk materials, the spin-spin correlation length ξ follows a critical behavior near the transition temperature: $\xi = \xi_0 \left[1 - \frac{T}{T_C(\infty)}\right]^{-\nu}$ where $\xi_0$ is a constant, ν is the critical exponent, $T_C(\infty)$ is the bulk transition temperature. At $T=T_C(\infty)$, ξ diverges so that any perturbation can trigger the magnetic transition, a characteristic of the continuous phase transition. For a finite system, the transition is smeared, and the effective transition is shifted towards the lower temperature since the size of the system limits the divergence of ξ. Equating ξ and $t-t_0$, the thickness of the optimal part of the film, the shifted transition temperature follows $1 - \frac{T_C(t)}{T_C(\infty)} = \left(\frac{\xi_0}{t-t_0}\right)^\nu$, which applies for $t-t_0 > \xi$. Fitting the data in Fig. 1(d) for $t > 2.7$ nm using the equation, one finds $T_C(\infty) = 420 \pm 10$ K, $\nu = 0.86 \pm 0.05$, and $\xi_0 = 1.8 \pm 0.1$ nm. The critical exponent ν is between those of the 2D and 3D Ising models [27], which is consistent with the strong perpendicular magnetic anisotropy of NCO (001) films [9].

### B. XMCD spectroscopy

To elucidate the nature of the interfacial layer, we measured XMCD spectra of NCO films with various thicknesses.

The magnetic hysteresis loop of a 14.2 nm NCO film was measured using the TFY mode, which measures the emitted photons by the excitation and subsequent relaxation of the electron bound to the measured cation. Therefore, the TFY mode has a larger probing depth in comparison to the TEY mode and the obtained XMCD spectra are more representative of the entirety of the film. As shown in Fig. 2(a), the square-shaped hysteresis loop measured at the Ni $L_3$ edge at 300 K agrees with that measured using magnetometry, as expected.

On the other hand, XMCD measurements reveal addition features in the hysteresis loop of the ultrathin films. As shown in Fig. 2(b), the magnetic hysteresis loop of a 1.6 nm NCO film was measured at 90 K using the TEY mode. The TEY mode, which probes only the top several nm of film portion where electrons can be liberated to the vacuum, is typically more suitable for ultrathin films because of the higher sensitivity. The hysteresis loops measured at the Co and Ni $L_3$ edges show opposite alignments with field, as is expected from the ferrimagnetic order of $NiCo_2O_4$ [10]. In addition to the sharp transition of the XMCD signal near the coercive field, changes are also observed near the zero field (see also Fig. S2) in the 1.6 nm film, but much less obvious in the thick films (see Fig. S3 and Fig. S4). For both the Co and Ni $L_3$ edges, when the magnetic field is ramped down from the high positive side, the XMCD signal drops substantially first near 100 Oe, which is before the magnetic field reverses the sign.

To reveal more details of the magnetic hysteresis at low fields for thin films, a magnetic hysteresis loop measured at 100 K from the TEY XMCD signal at the Ni $L_3$ edge for the 1.6 nm film is plotted in Fig. 3(a) where Features 1 and 2 are defined. Temperature dependence of hysteresis loops from this sample is shown in Fig. 3(b). At 100 K (below $T_C$), Feature 1 remains prominent. As temperature increases, the magnitude of Feature 1 reduces. Plotting temperature dependence of the magnitude of Feature 1 along with the magnetization measured using magnetometry shows remarkable agreement [Fig. 3(c)], demonstrating that Feature 1 constitutes the magnetic transition in the 1.6 nm film observed by magnetometry. In contrast, the magnitude of Feature 2 is still substantial at 150 K and remains nonzero at 300 K.

The shape of Feature 1 in Fig. 3 (a) is biased differently on each side of zero-field, unlike a traditional exchange-biased system, switching before reaching zero-field. Such behavior is similar to that seen in exchange-spring systems [29,30]. Such exchange-spring systems can be modeled as a bilayer magnetic system with antiferromagnetic coupling (denoted as major and minor components). The minor component experiences an effective field due to both applied and exchange fields, with exchange field anti-aligned with the major component. An illustration of the exchange-spring switching pattern described is shown in Fig. 3(d), with correspondence to the hysteresis loop labeled. At high applied field, the effective field is sufficient to align both components [(1) in Fig. 3(d)]. As the field approaches zero, the minor component is allowed to become anti-aligned with the rest of the film, resulting in a decrease of the measured XMCD signal [(2) and (3) in Fig. 3(d)]. Once the field applied is sufficient to switch the major component of the film, the entire film becomes magnetized again with the external field [(4) in Fig. 3(d)].

*C. SPIPES measurement*

To probe the magnetism of the surface layer, we measured the SPIPES spectra of a 12 nm NCO films. In the SPIPES measurement, a pulsed magnetic field was applied in-plane and the spectra were taken at remanence with electrons which were spin polarized in-plane, as described elsewhere [17]. SPIPES spectra can be compared to XAS and XMCD spectra since each of the techniques probes unoccupied states above Fermi level [31,32]. The SPIPES spectrum in Fig. 4(a) contains spin majority, spin minority and spin integrated density of states. The spin majority (minority) density of states is obtained when the spin polarization of incident electron in plane of the sample is parallel (opposite) to the applied magnetic field. A noticeable difference in spin majority and spin minority intensity was observed in the unoccupied states above Fermi level indicating spin polarized unoccupied states of the NCO thin films. In-plane polarization can be estimated using the difference of data between spin majority and spin minority density of states. Two appreciable in-plane polarization values of 4.7 % and 4.0 % were estimated in the unoccupied states at 2.2 and 4.2 eV above Fermi level respectively. The separation of 2.0 eV between the polarized states in SPIPES is qualitatively in good agreement with the separation of about 1.9 eV between two features in XMCD spectra of both Co and Ni as shown in Figs. 4(b) and (c). The comparison shows that the states in SPIPES are shifted to higher energies. This can be ascribed to different final state effects in SPIPES and XMCD, inducing some shifts in apparent binding energies or photon energies [31,33]. Although the in-plane polarization is small, the value is non-zero, which indicates that the magnetic moments have a surface component, and the moments therefore are slightly canted in surface region. Furthermore, comparison of SPIPES and hysteresis loops obtained by XMCD shows that Co (Ni) is spin minority (majority) carrier, and features above Fermi level is mostly Co weighted.

*D. Discussion*

The exchange-spring behavior has been observed previously in NCO films grown in low $P_{O2}$, which show separated regions of optimal NCO and $T_d$-vacant NCO [14,34,35]. The lateral dimension of these separated regions is on the order of several nanometers, consistent with our interfacial layer thickness calculation above. Essentially, at low $P_{O2}$, the population of the $Ni^{3+}$ on the $O_h$ decreases due to the reduced stability [12]. These $O_h$ sites are then filled by Co ions, generating $T_d$ site vacancies. In optimal NCO, the total magnetic moment is parallel to the moments on the $T_d$ sites and antiparallel to those on the $O_h$ sites according to the ferrimagnetic order [12]. In contrast, in the $T_d$-vacant NCO, the total moment can be parallel to the moments on the $O_h$ sites if the occupancy of the $T_d$ sites is low enough. Given the antiferromagnetic exchange interaction between the $O_h$ and the $T_d$ sites, the magnetization of the adjacent optimal NCO and $T_d$-vacant NCO is expected to have antiferromagnetic coupling, leading to the exchange-spring behavior.

A similar scenario can be pictured in the interfacial layer in which the cation stoichiometry may deviate from the optimal values. Previous work has demonstrated diffusion of Mg into inverse spinel epitaxial $Fe_3O_4$ films grown on MgO substrates when the substrate temperature reaches 350 °C [36]. Moreover, the diffusion favors a vacancy mechanism, suggesting that Mg ions tend to occupy the cations sites. The random replacement of Ni and Co ions with non-magnetic $Mg^{2+}$ ions can generate magnetic frustration because of the coexisting ferromagnetic and antiferromagnetic exchange interactions in NCO [12], which manifests in spin-glass like behavior featuring measurement-speed-dependent magnetic properties. As shown in Fig. 5, magnetization-temperature relation was measured with different speed. The faster (3 K/min) measurement was the same as that displayed in Fig. 1(b), which shows a clear magnetic transition at about 120 K. The slower (1 K/min) measurement, however, uncovers another component that remains magnetic above 120 K, a behavior similar to that of Feature 2 in Fig. 3(a). The slow magnetic response of this component is typical for the spin glass states.

**Conclusions**

In conclusion, we have presented evidence that the $NiCo_2O_4$(001)/$MgAl_2O_4$ (001) films consist of a layer of optimal stoichiometry and an interfacial layer. The behavior of the optimal layer can be understood using the finite scaling theory with transition temperature $T_C(\infty) = 420 \pm$ 10 K, spin-spin correlation length $\xi_0 = 1.8 \pm 0.1$ nm, and critical component $\nu = 0.86 \pm 0.05$ which is consistent with the high perpendicular anisotropy. The interfacial layer couples anti-ferromagnetically to the optimal layer, indicating its deviation from optimal stoichiometry, possibly due to the interfacial diffusion. Non-zero in-plane polarization of the unoccupied states above Fermi level was observed, indicating that magnetic moments are canted near the surface region. These results reveal fundamental parameters of NCO magnetism as well as the possible disordered nature of the NCO/MAO interface which is critical for the application of ultrathin films.

**Acknowledgement**: The authors acknowledge the primary support from the National Science Foundation (NSF) through EPSCoR RII Track-1: Emergent Quantum Materials and Technologies (EQUATE), Award No. OIA-2044049 (UNL) and through DMR #1708790 (BMC). This research used resources of the Advanced Photon Source, a U.S. Department of Energy (DOE) Office of Science User Facility operated for the DOE Office of Science by Argonne National Laboratory under Contract No. DE-AC02-06CH11357. This research was performed in part at the Nebraska Nanoscale Facility, National Nanotechnology Coordinated Infrastructure and the Nebraska Center for Materials and Nanoscience, which are supported by the NSF under grant no. ECCS- 2025298, as well as the Nebraska Research Initiative through the Nebraska Center for Materials and Nanoscience and the Nanoengineering Research Core Facility at the University of Nebraska–Lincoln.

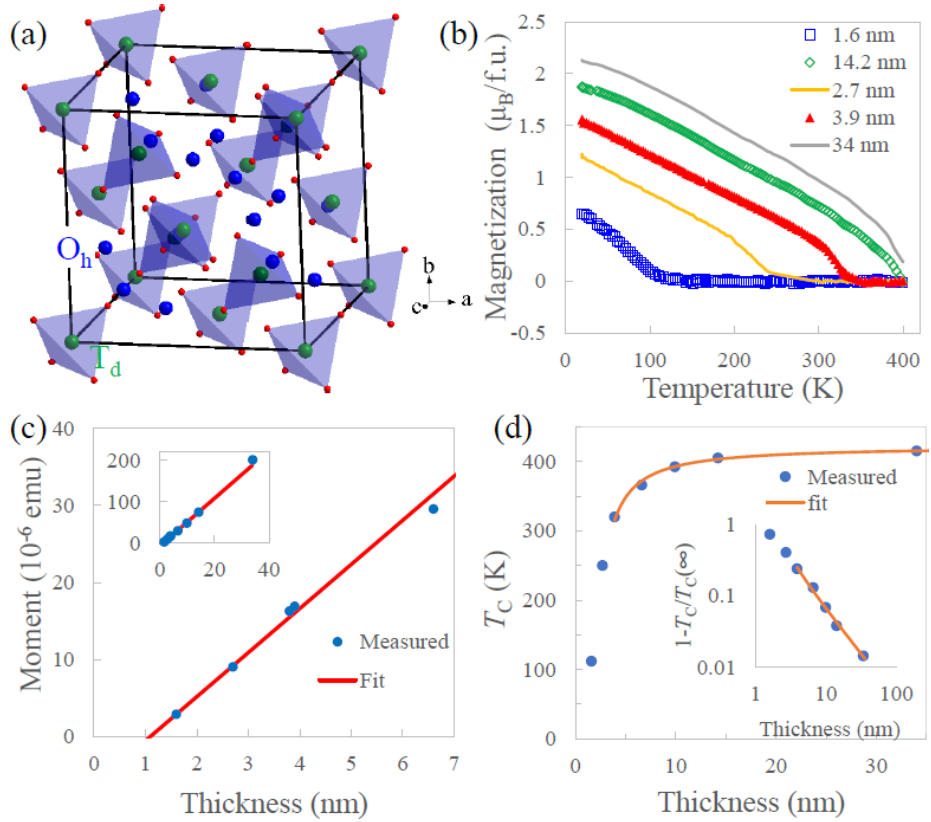

**Figure 1** (a) Crystal structure of NCO where the $T_d$ and $O_h$ sites are indicated. (b) The temperature dependence of magnetization of the films with various thicknesses measured by magnetometry at 200 Oe. (c) Total magnetic moment at 20 K as a function of thickness and the linear fit. (d) Magnetic transition temperature $T_C$ as a function of thickness and the fit using the finite-scaling theory.

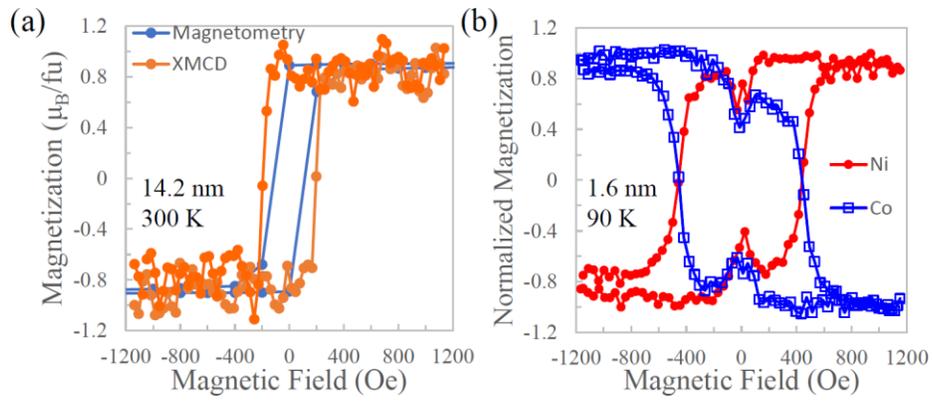

**Figure 2**. Magnetic hysteresis loops. (a) Comparison between the measurements of the 14.2 nm at 300 K film using magnetometry and XMCD. (b) Element-specific loops for the 1.6 nm film at 90 K.

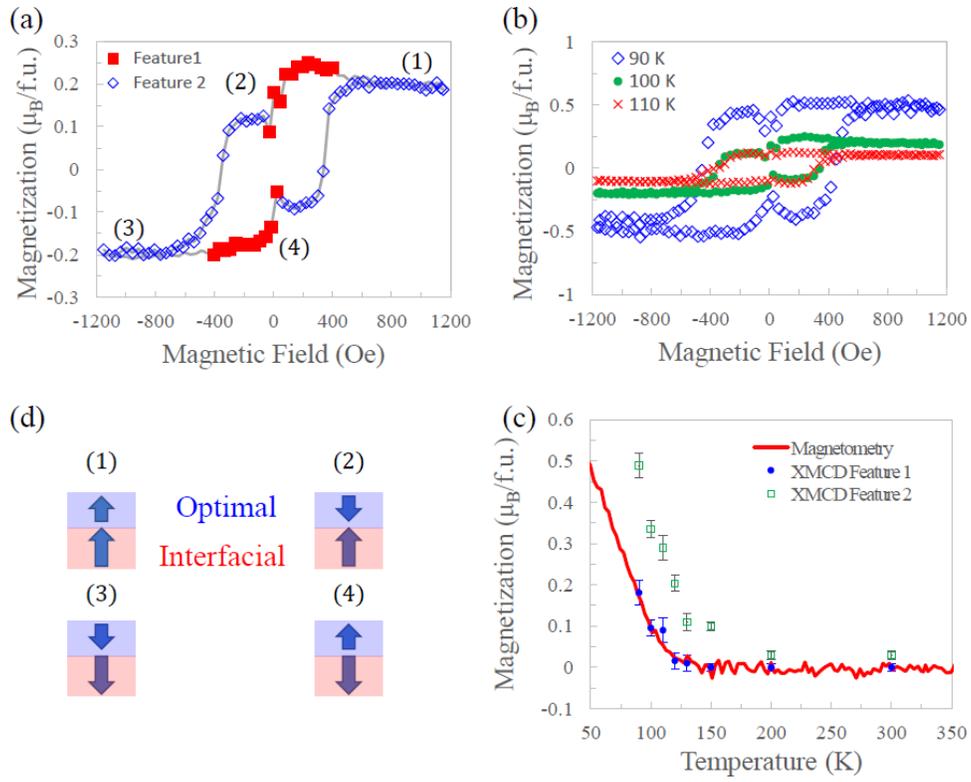

**Figure 3**. Magnetic properties of the 1.6 nm film. (a) Hysteresis loops measured using Ni L$_3$ edge at 100 K with two features identified. (b) Hysteresis loops measured at different temperatures. (c) The temperature dependence of the magnetization of the two components and the comparison with the magnetometry measurement. (d) Schematic illustration of the exchange spring behavior with two magnetic components of antiferromagnetic coupling.

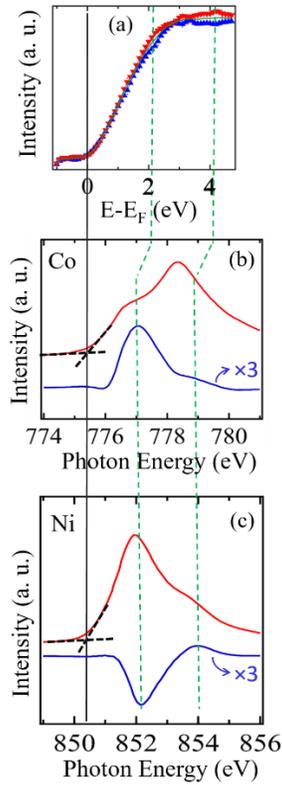

**Figure 4**. The comparison of the SPIPES, XMCD, and XAS spectra for a 12 nm $NiCo_2O_4$ films. (a) The SPIPES spectrum with the spin majority (blue upright triangles), spin minority (red inverted triangles), and spin integrated (green solid line) density of states. The XAS (red solid lines) and XMCD (blue solid lines) spectra for (b) Co and (c) Ni $L_3$ edges. The black vertical line aligns the Fermi level, and green dashed lines indicate the possible correspondence between unoccupied states in SPIPES, XAS and XMCD spectra. The XMCD spectra have been magnified by three times for better visualization.

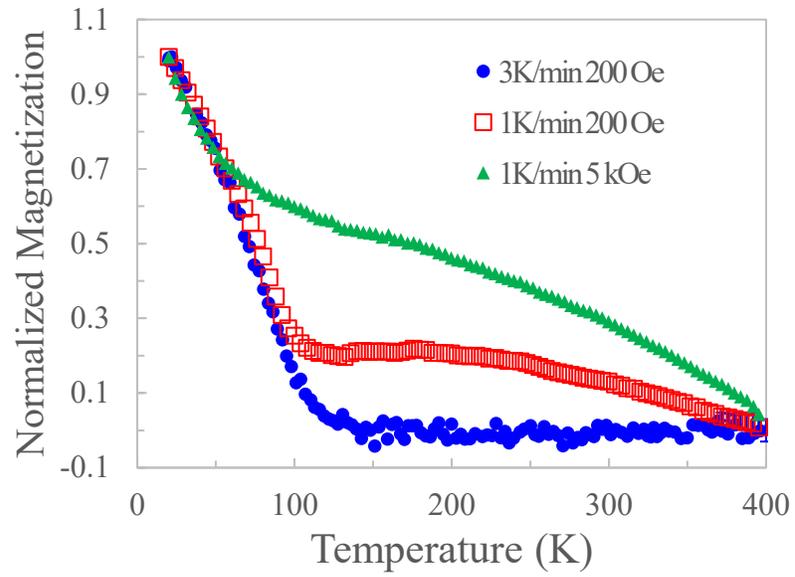

**Figure 5**. Magnetization-temperature relation of the 1.6 nm NCO film measured by magnetometry with different speed and magnetic field.